\documentclass[oribibl,envcountsame]{llncs}  % use original bibliography and make theorem-like environments use the same counter

\usepackage{epsfig}
\usepackage{tikz}
\newcommand{\ignoreText}[1]{}
\usepackage{multirow}
\usepackage{rotating}
\usepackage{array}
\usepackage{subfigure}

\usetikzlibrary{shapes,snakes}

\title{Building conceptual spaces for exploring and linking biomedical resources}

\author{%
  R. Berlanga \and
  E. Jim\'enez-Ruiz \and %inst{1}
  V. Nebot
}
\institute{%
  Departamento de Lenguajes y Sistemas Inform\'aticos \\
  Universitat Jaume I, Spain, \\
  \url|{berlanga,ejimenez,romerom}@uji.es|
}

\begin{document}

\maketitle

\begin{abstract}
The establishment of links between data (e.g., patient records) and Web resources (e.g., literature) and the proper
visualization of such discovered knowledge is still a challenge in most Life Science domains (e.g., biomedicine). In this paper
we present our contribution to the community in the form of an infrastructure to annotate information resources,
to discover relationships among them, and to represent and visualize the new discovered knowledge. Furthermore, we
have also implemented a Web-based prototype tool which integrates the proposed infrastructure.
\end{abstract}

\section{Introduction}

The ever increasing volume of web resources as well as generated data from automated applications is challenging current approaches for biomedical information processing and analysis. One current trend is to build semantic spaces where corporate data and knowledge resources can be mapped in order to ease their exploration and integration. Semantic spaces are usually defined in terms of widely accepted knowledge resources (e.g. thesauri and domain ontologies), and they are populated by applying (semi)automatic semantic annotation processes.

Apart from these semantic spaces it is also crucial to propose new summarization tools that help both users and machines to better analyze and extract knowledge
from these spaces. On-Line Analytical Processing (OLAP) techniques have been very successfully used to analyze summarized data from different perspectives (dimensions) and detail levels (categories). However, OLAP cannot be directly applied to the aforementioned semantic spaces for several reasons: first, data resources and knowledge are highly heterogeneous and  dynamic and second, semantic annotations are based on graph structures which make it difficult their translation to OLAP multidimensional spaces. Despite these limitations, OLAP operators could be very useful as they provide an intuitive and interactive way to explore multidimensional spaces.

In this paper we propose a new visual paradigm, called 3D conceptual maps, which allows users to explore and analyze interesting associations derived from data and web resources, which have been previously annotated with a reference domain ontology. Conceptual maps can be dynamically built according to the users analysis requirements, and they provide interactivity through operators similar to traditional OLAP operators (e.g. drill-down, roll-up, etc.) The main novelty 
of the new operators is that they are semantic-aware, that is, they take into account the semantics of the domain ontologies to summarize the data that is visualized in the conceptual maps. We also present a web-based prototype tool called \textit{3D knowledge browser} (3DKB), which integrates the previous visual paradigm and operations.

As far as we know, there are no similar tools in the literature which allow summarizing and exploring discovered concepts and relationships from different biomedical
sources (not only literature). Previous work exists on discovering biomedical relationships from semantic annotations, for example \cite{EBIMed2007,MedEvi2008,FACTA2008} to mention a few, but they are limited to present results as tabular data, and the target collection is always PubMed abstracts. Instead, our proposal is aimed to deal with multiple sources (e.g. PubMed abstracts, patient records, public databases, and so on) and it provides mechanisms to explore the discovered relationships through the reference ontologies.

The paper is organized as follows. In Section 2 we introduce the motivating scenario. Then,  Sections 3 and 4 present our prototype and its use through two use cases. Section 5 is devoted to the methodological aspects.   First, we describe the normalization formalism to represent both the knowledge resources and the target collections. Then, we introduce the main operators required over the normalized representation to provide interactivity with the conceptual maps. Finally, we give some conclusions and future work.

\section{Motivating Scenario}

The need of semantically integrating different biomedical sources arose in the context of the
European Health-e-Child (HeC) \cite{HeC06,HeC06_2} integrated project. HeC aimed to develop an 
integrated health care platform to allow European paediatrics to access, 
analyse, evaluate, enhance and exchange integrated biomedical information focused 
on three paediatric diseases: (1) heart disorders, (2) inflammatory disorders and (3) brain tumours. 
The biomedical information sources covered six distinct levels of granularity (also
referred to as vertical levels), classified as molecular (e.g., genomic and
proteomic data), cellular (e.g., results of blood tests), tissue (e.g.,
synovial fluid tests), organ (e.g., affected joints, heart description),
individual (e.g., examinations, treatments), and population (e.g.,
epidemiological studies).

The 3DKB tool is mainly aimed at providing an
integrated and interactive way to browse biomedical concepts as well as to
access external information (e.g., PubMed abstracts) and HeC patient data
related to those concepts. The 3DKB is intended to facilitate the integration by providing the
clinician with a predefined subset of semantically annotated web objects that
are relevant to her domain. These objects are thus implicitly linked to
clinician and patient data, which are also semantically
annotated with the same knowledge resource.

In our current implementation, we selected the Unified Medical Language System
Metathesaurus (UMLS) \cite{UMLS2004} as the knowledge resource with which semantic
annotations are generated. UMLS represents the main effort for the
creation of a multipurpose \textit{reference thesaurus}. UMLS contains
concepts from more than one hundred terminologies, classifications, and
thesauri; e.g. FMA, MeSH, SNOMED CT or ICD. UMLS includes two million terms and
more than three million term names, hypernymy classification with more than one million
relationships, and around forty millions of other kinds of relationships.

\section{Prototype Implementation}

\begin{figure}[t]
 \centering
	 \includegraphics[width=0.9\textwidth]{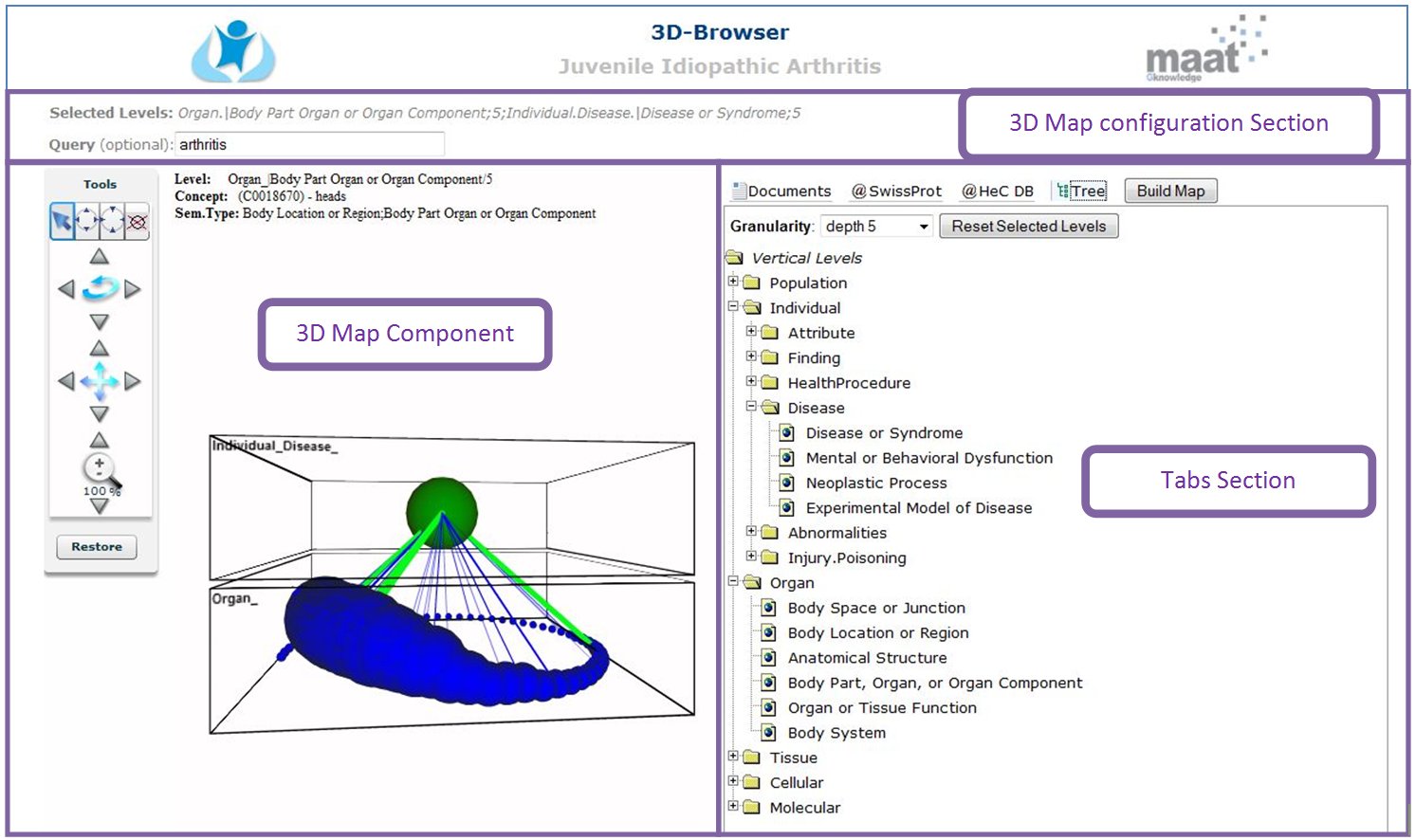}
  \caption{3D Knowledge Browser snapshot with its main visual components}
  \label{fig:3dbrowser}
\end{figure}

The current prototype has been developed using AJAX (Asynchronous JavaScript
and XML) technologies. Figure \ref{fig:3dbrowser} shows an overall view of the
3DKB tool \cite{3dbrowser} for the JIA domain. It consists of three main parts,
namely: 1) the configuration of the 3D Conceptual Map (from now on 3D-Map), which contains the selected
vertical levels (i.e., HeC levels) and an optional free text query to evaluate
against the visualized concepts, 2) the 3D-Map itself, which contains the
biomedical concepts stratified in vertical levels according to the previous
configuration, and 3) a series of tabs that contain a ranked list of objects
associated to a selected concept from the 3D-Map. In the latter, each tab
represents a different type of object (e.g., PubMed abstract, Swissprot protein
and HeC patient data). There is a special tab entitled ``Tree'' which contains
all the possible levels that can be selected to configure and build the 3D-Map.
The levels are based on the UMLS semantic types
\cite{semnetwork1989,SemGroupsBodenreider2003} which are grouped within the
correspondent HeC levels as in \cite{umlstagger07,BMC08}. 
The layers of the 3D-Map can be defined by selecting levels of the ``Tree'' tab and also through a keyword-based query. In the second case, only the most specific concepts whose lexical forms match the query are visualized.

The visual paradigm of 3D-Maps relies on the vertical integration vision 
proposed in HeC. That is, all the involved knowledge,
data and information are organized into different disjoint conceptual levels
(i.e., vertical levels), each one representing a different perspective of the
biomedical research. In this
way, the 3DKB presents a stratified view of the information based on
vertical levels (see \emph{Individual.Disease} and \emph{Organ boxes} in
3D-Map of Figure \ref{fig:3dbrowser}). Within each level, biomedical concepts 
deemed relevant for both the clinician domain (e.g., rheumatology, cardiology and
oncology) and the clinician information requests are shown as \emph{balls} in
the 3D-Map. Relevance of concepts is defined in terms of the collection
frequency (e.g., PubMed abstracts), and it is represented in the 3D-Map through
the ball size. Regarding the color of the ball, normal concepts are displayed
in blue, expanded concepts in red and concepts containing query entities in green.

\emph{Semantic bridges} are another important visual element of the 3D-Map,
which are defined as links between concepts of two different vertical levels
and they are represented as 3D lines in the 3D-Map. Semantic bridges can
represent either co-occurrences of concepts in the target collection or well-known 
relationships between concepts stated in some domain ontology (e.g.,
 UMLS). Semantic bridges can help clinicians to select the context in which the
 required information must hold. For example, from the 3D-Map in Figure
 \ref{fig:3dbrowser} we can retrieve documents or patient IDs about 
 \emph{arthritis related to limb joints} by clicking an existing
 bridge between the concepts \emph{Arthritis} and \emph{Limb\_Joints}. Finally,
 semantic bridges have also associated a relevance index, which depends on the
 correlation measure we have chosen for their definition (e.g. count, log
 ratio, odds ratio, etc.). 

\begin{figure*}[t]
\centering{
	%\fbox{
	\subfigure[]{\label{fig:3dexpandA}
    \includegraphics[width=0.47\textwidth]{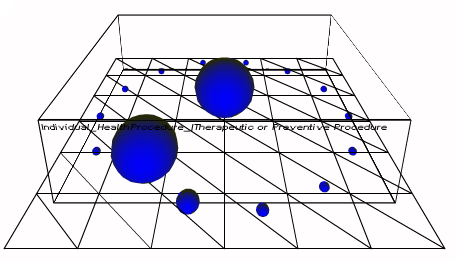}
    %}
    } 
    %\fbox{
    \subfigure[]{\label{fig:3dexpandB}
    \includegraphics[width=0.47\textwidth]{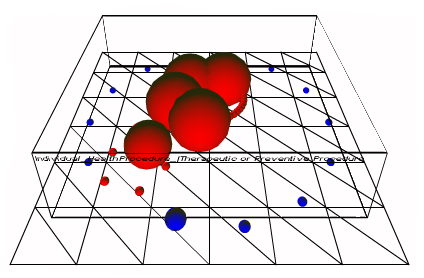}
    }
    %}
\caption{Example of two expanded concepts: Operation and Implantation}
\label{fig:3dexpand}
}
\end{figure*}

Another interesting feature of 3D-Maps is the ability of browsing through the
taxonomical hierarchies of the biomedical concepts (e.g., UMLS hierarchy). In
the example of Subfigure \ref{fig:3dexpand}, the user can expand the concepts
\emph{Operation} and \emph{Implantation} (biggest balls in Figure
\ref{fig:3dexpandA}). The resulting concepts are
red-coloured (Subfigure \ref{fig:3dexpandB}) and represent more specific
concepts like \emph{Catheterisation}, \emph{Surgical repair}, \emph{Intubation},
or \emph{Cardiovascular Operations}.

In order to manage the elements of the 3D-Map a series of operations are
provided in the 3D-Map tools panel (see left hand-side of Figure
\ref{fig:3dbrowser}). These operations are split within two categories:
operations to manage the whole 3D-Map (rotate, zoom and shift) and
concept-related operations. The operations to manage the concept visualization
involve (1) the retrieval of the objects associated to the clicked concept, (2)
the expansion of the clicked concept, (3) the removal of the concepts of
a level with the exception of the clicked concept, and (4) the deletion of the
clicked concept.

\section{Use Cases}

In this section we will show the functionalities of the 3DKB through two
use cases based on some HeC clinician information requests.

\subsection{Case 1: Exploring the relation between procedures and results in the
Tertalogy of Fallot (ToF) domain}

\begin{figure*}[t]
\centering{
	%\fbox{
	\subfigure[]{\label{fig:case1a}
    \includegraphics[width=0.47\textwidth]{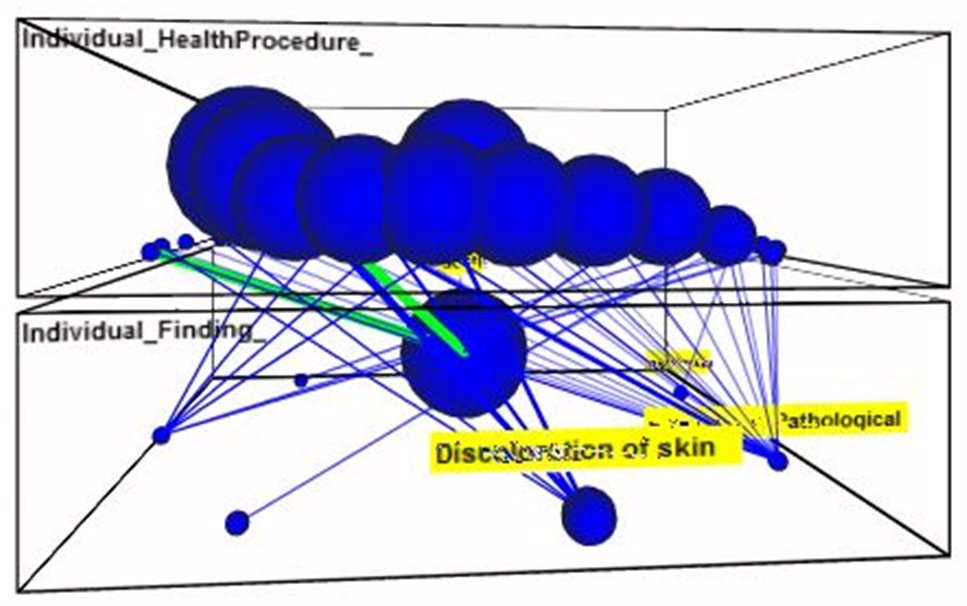}
    %}
    } 
    %\fbox{
    \subfigure[]{\label{fig:case1b}
    \includegraphics[width=0.47\textwidth]{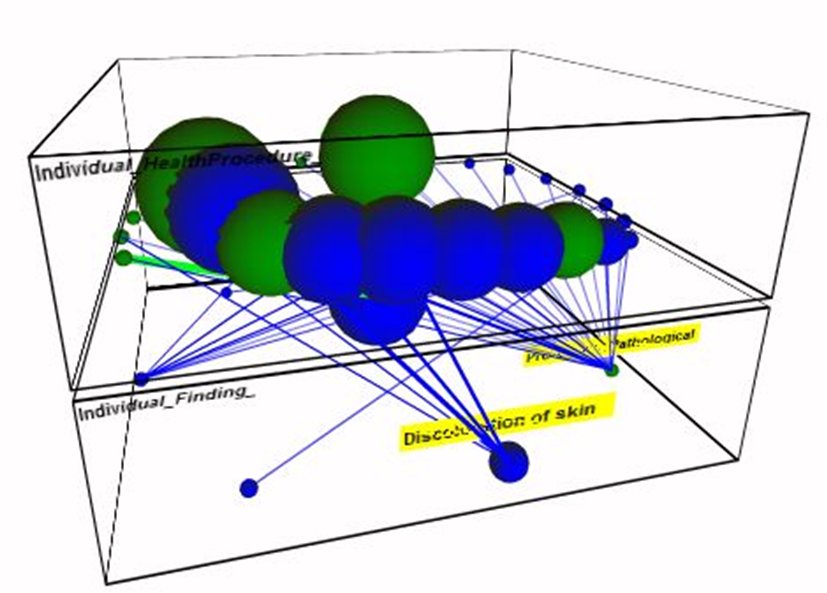}
    }
    
    \subfigure[]{\label{fig:case1c}
    \includegraphics[width=0.47\textwidth]{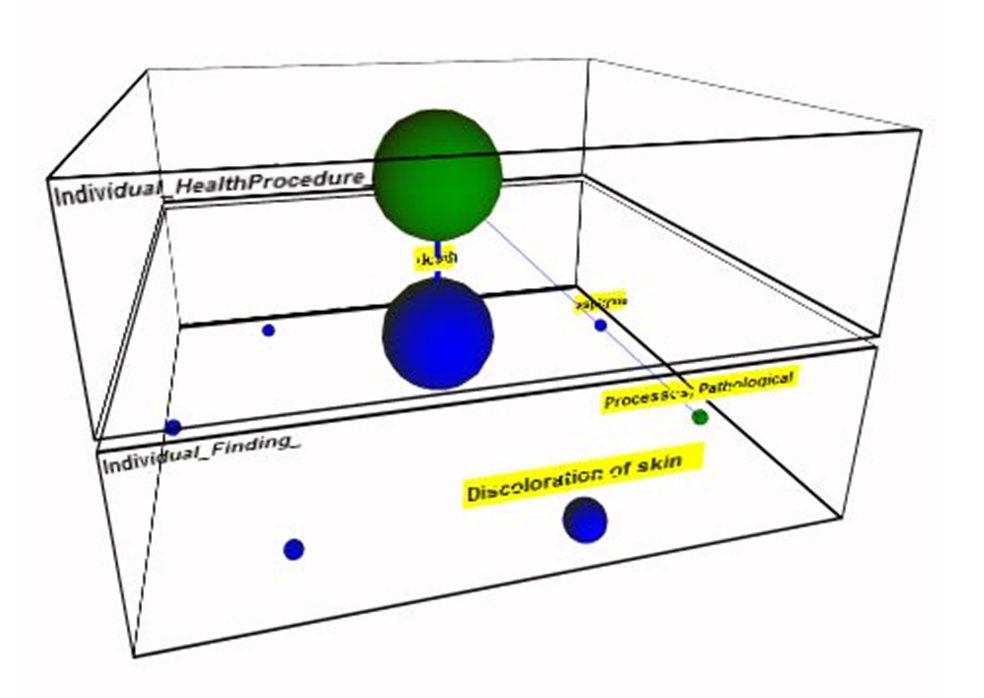}
    %}
    } 
    %\fbox{
    \subfigure[]{\label{fig:case1d}
    \includegraphics[width=0.47\textwidth]{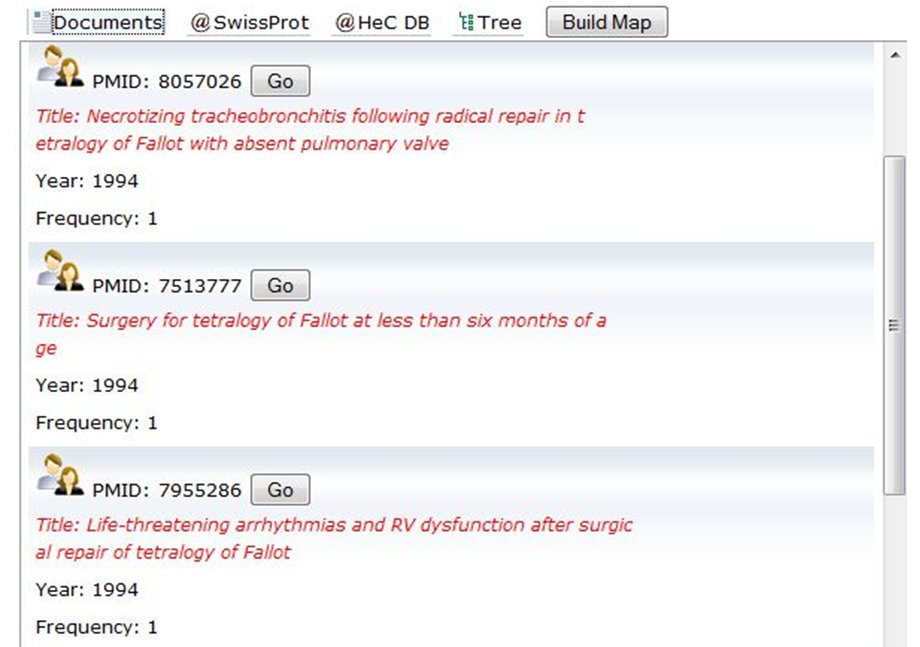}
    }
    %}
\caption{Interesting relationships between procedures and findings in the
literature}
\label{fig:case1}
}
\end{figure*}

In this case, the clinician is interested in knowing the relation between the
different surgical techniques reported in the literature and the findings and
results that are usually correlated to them. For this purpose, the clinician
builds the 3D-Map for the semantic levels
\emph{Individual.Health\_Procedures.} and
\emph{Individual.Finding}.
As a result the clinician obtains the map presented in Figure
\ref{fig:case1a}.  However, the clinician is only interested in
repair operations. So, she refines the query by specifying the keyword �repair�
in the query input field. The resulting 3D-Map is shown in the Figure
\ref{fig:case1b}, where relevant concepts are coloured in green. These relevant
concepts contain at least one sub-concept (including itself) matching the specified query. 
Now, the clinician can select one of the green-coloured concepts, for
example \emph{Repair Fallot Tetralogy}, in order to filter the map to just those
concepts that are related to it (see Figure \ref{fig:case1c}).
Finally, she finds an interesting bridge between the selected concept and the
finding concept \emph{Death}. Figure \ref{fig:case1d} shows the
documents that are retrieved by clicking this bridge. Notice that these
abstracts are about death cases related to TOF repair.

\subsection{Case 2: Finding potential proteins that can be related to different
types of a disease within the Brain Tumours (BT) domain}

\begin{figure}[t]
 \centering
	 \subfigure[]{\label{fig:case2a}
    \includegraphics[width=0.47\textwidth]{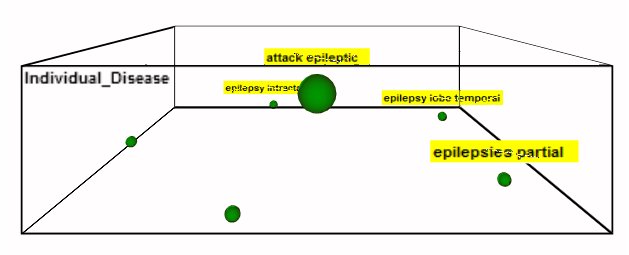}
    %}
    } 
    %\fbox{
    \subfigure[]{\label{fig:case2b}
    \includegraphics[width=0.47\textwidth]{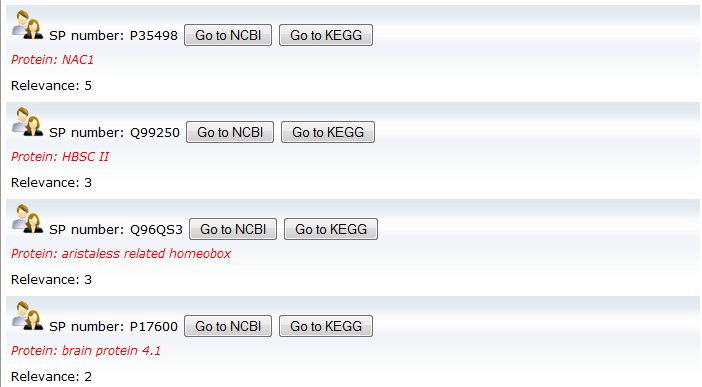}
    }

  \caption{Proteins retrieved through the @Swissprot tab for concept
  \emph{attack epileptic}}
  \label{fig:case2}
\end{figure}

In this use case, the clinician is interested in comparing the proteins related
to a disease and its subtypes. Taking the brain tumour domain, the clinician
specifies the concept query \emph{epilepsy} without selecting any vertical
level. As a result, she obtains the 3D-Map of Figure \ref{fig:case2a} which
contains the concepts \emph{attack epileptic}, \emph{epilepsy intractable},
\emph{epilepsy lobe temporal}, \emph{epilepsy extratemporal} and \emph{epilepsy
focal}.

To retrieve the proteins related to these diseases, the tab �@SwissProt� is
selected. For example in Figure \ref{fig:case2b}  the related proteins
to \emph{attack epileptic} are shown. The user can then get much more information about
these  proteins by clicking the buttons �NCBI� and �KEGG�, which jump to the
corresponding pages in Entrez Gene and KEGG sites respectively.
Note that, the relevance of each protein entry is
calculated with the frequency of the concept and its sub-concepts in the
Swissprot DB description of the protein.

%Notice that selected concepts have overlapping sets of associated proteins
%but also other different proteins.

\section{Method}

OLAP (On-line Analytical Processing) \cite{codd1993} tools were introduced
to ease information analysis and navigation from large amounts of transactional data.
OLAP systems rely on multidimensional data models, which are based on the fact/dimension dichotomy. 
Data are represented as facts  (i.e. subject of analysis), while dimensions contain a hierarchy of levels, which provide different granularities to aggregate the data. One fact and several dimensions to analyze it
give rise to what is known as the data cube. Common operations include \emph{slice} (i.e. performing a selection on one dimension of the given cube, thus resulting in a sub-cube), \emph{dice} (i.e. similar to slice but  performing a selection on two or more dimensions), \emph{drill-down} (i.e. navigating among levels of data ranging from the most summarized (up) to the most detailed (down)), \emph{roll-up} (i.e. inverse of drill-down, that is, climbing up the concept hierarchy) and \emph{pivot} (i.e. rotate the data to provide an alternative presentation).

Since multidimensionality provides a friendly, easy-to-understand and intuitive visualization of data for non-expert end-users, we have borrowed the previous concepts and operations to apply them to our 3D conceptual maps.

\subsection{Representation of Semantic Spaces}
In order to achieve a browsable analytical semantic space, it is necessary to normalize the representation of both the knowledge resource and the target collection (e.g., patient records, PubMed abstracts, and so on).
This normalization consists of two main steps: (1) to arrange existing concepts into a well-structured multidimensional schema, and (2) to represent the objects collection under this schema. The first step must be guided by a series of predefined \emph{dimensions} which roughly represent \emph{semantic groups}. For example,
in the HeC project dimensions correspond to vertical levels: population, disease, organ, and so on. The main issue to be addressed in this step is the irregular structures of the taxonomies provided
by existing knowledge resources. The second step has two main tasks: (1) to semantically annotate the objects collection  with concepts from the knowledge resource,
and (2) normalize the annotation sets of each object to the multidimensional schema defined in the previous step. The subsequent sections are devoted to 
describe all this process in detail.

\subsubsection{Semantic Annotation}
During the last years, we have witnessed a great interest in massively annotating biomedical scientific literature. Most of the current annotators rely on well-known lexical/ontological resources such as MeSH, Uniprot, UMLS and so on. These knowledge resources usually provide both the lexical variants for each inventory \emph{concept} and the concept taxonomies. Some knowledge resources are more formal (e.g. FMA, Galen, etc.), providing logic definitions for concepts from which the taxonomy can be inferred.

%Problem statement
In our work, the knowledge resource used to generate semantic annotations is called  \emph{reference ontology}, denoted $O$. The lexical variants associated to each ontology concept $c$ is denoted with $lex(c)$, which is a list of strings. The taxonomic relations between two concepts $a$ and $b$ is represented as $a \preceq b$. A semantic annotation of a text fragment $T$ consists of identifying the concepts in $O$ such that they are more likely to represent the meaning of $T$.

Most semantic annotation systems are dictionary look-up approaches, that is, they rely on the lexicon provided by the ontology in order to map text words to concept lexical variants. Some popular annotation systems in the biomedical domain are Whatizit \cite{Whatizit2008} and MetaMap \cite{Aronson2001}. Current research of semantic annotation is focusing on scalability issues, and the definition of gold \cite{GENIA} and silver standards \cite{CALBC} to evaluate the quality of these systems. In these standards, an XML format IeXML has been proposed to represent the generated semantic annotations. An example of this format is shown in Figure \ref{taggedtext}, which was generated with our annotation system.

\begin{figure}
{\scriptsize
\begin{verbatim}
<e id="UMLS:C1709323:T062::1,2"><w id="1">Open</w> <w id="2">label</w></e> 
<e id="UMLS:C0282460:T062::1,2,3"><w id="1">phase</w> <w id="2">II</w><w id="3">trial</w></e> 
of <e id="UMLS:C0205171:T081">single</e>, <e id="UMLS:C0205385:T080">ascending</e> 
<e id="UMLS:C0439568:T079">doses</e> of MRA in  
<e id="UMLS:C0007457:T098|UMLS:C0043157:T098">Caucasian</e><e id="UMLS:C0008059:T100">children</e> 
with <e id="UMLS:C0205082:T080">severe</e> 
<e id="UMLS:C1384600:T047::1,2,3,4|UMLS:C0682057:T100::2"><w id="1">systemic</w> 
<w id="2">juvenile</w><w id="3">idiopathic</w><w id="4">arthritis</w></e>: proof of principle of 
the <e id="UMLS:C1707887:T062">efficacy</e> of 
<e id="UMLS:C0063717:T116,T129,T192::1,2"><w id="1">IL-6</w> <w id="2">receptor</w> </e> 
<e id="UMLS:C0332206:T169">blockade</e> in this 
<e id="UMLS:C0332307:T080|UMLS:C0455704:T170">type</e> of arthritis and demonstration of 
<e id="UMLS:C0439590:T079">prolonged</e> <e id="UMLS:C0205210:T080">clinical</e> improvement.</s>
\end{verbatim}
}
\caption{Example of tagged text with IeXML.}\label{taggedtext}
\end{figure}

One of the main drawbacks of current semantic annotation systems is that they usually focus on very specific entity types like proteins and diseases. In our work, we aim to generate semantic annotations of any entity type involved in the biomedical research. For this reason, we have chosen the UMLS-Meta as knowledge resource, which provides more than 100 entity types (semantic types). However, just a few annotation systems are able to manage the huge amount of lexical information provided by UMLS-Meta, and they are too slow to deal with large text collections. As a consequence we developed a novel annotation system, called Concept Retrieval, which is based on information retrieval techniques to efficiently perform the text annotation \cite{BioAnnotator2010}. This annotation system was tested in the CALBC competition over a collection of 150.000 PubMed abstracts about immunology \cite{CALBC}.

\subsubsection{Knowledge normalization}

In order to build semantic spaces for analyzing document collections, the reference ontology $O$ associated to the knowledge resource is normalized as follows:

\begin{itemize}
\item First a set of dimensions are defined, $(D_1, \cdots D_n)$, which represent a partition of the concepts in the domain ontology. 
Each dimension $D_i$ represents a different semantic space (e.g. semantic types or vertical levels), and cannot share any common sub-concept with the other dimensions. 

\item Each dimension $D_i$ can define a set of categories or levels $L^i_j$, which forms in turn a partition over $D_i$ but with the following constraints:
(1) there cannot be two concepts $c$ and $d$ in $L^i_j$ such that either $c \preceq d$ or $d \preceq c$, and (2) all the concepts in $L^i_j$ have a common super-concept that belongs to $D_i$.

\item Every concept of the ontology is encoded under the labeling scheme presented in \cite{INS09}. Thus, each concept $c \in O$ is represented with the following descriptor:

 \[ \langle c, pre\_index, anc\_index, desc\_intervals, anc\_intervals, topo\_order \rangle \]
 
where $pre\_index$ is the pre-order index in the spanning tree of $O$, $desc\_intervals$ is the list of index intervals of the descendants of $c$ (i.e., $\{c'|c' \preceq c\}$), 
$anc\_index$ is the pre-order index of the reversed spanning tree, and $anc\_intervals$ is the list of index intervals of the ancestors of $c$. Finally, $topo\_order$ is the topological order of the concept in the spanning tree of $O$. More specifically, this descriptor represents two labeling schemes, namely: $\mathcal{L}^-$ for descendants, and $\mathcal{L}^+$ for ancestors.
Under these labeling schemes, queries over the taxonomical relationships are efficiently computed with a specific interval algebra \cite{INS09}.
\end{itemize}

One interesting application of the labeling scheme $\mathcal{L}^+$ is the efficient construction of ontology fragments tailored to an input set of concepts, called signature.
In this way, we can automatically build each dimension $D_i$ with the ontology fragment obtained with the signature formed by all the concepts identified in the collection (through
semantic annotation) and that belong to some semantic group representing the dimension (e.g. disease, protein, and so on). To obtain the categories of a dimension $D_i$, we
take into consideration the taxonomic relationships in the fragment and the previous restrictions over dimensions and their categories.

\subsubsection{Data and resource normalization}

After semantic annotation, each document of the target collection ${\mathcal Col}$ has associated a list of concepts from the reference ontology $O$. However, these annotation sets are not suited for multidimensional analysis, and therefore a normalization process similar to that applied to the ontology must be performed. The main goal of objects normalization is to represent the semantic annotations within the normalized multidimensional space. Thus, each document $d \in {\mathcal Col}$ is represented as the multidimensional fact: 

         \[ fact(d) = (D_1=c_1,\cdots, D_n=c_n) \]

where $c_i$ $(0 \leq i \leq n)$ is either a concept from the dimension $D_i$ or the $null$ value. 
Remember that concepts are represented under the labeling scheme $\mathcal{L}^-$, and consequently they are expressed through their $pre\_index$ numbers.

As a semantic annotator can tag more than one concept of the same dimension, the normalization process consists in selecting the most relevant concepts for each dimension.
For this purpose, for each document $d$ we first build a concept affinity matrix $M^d$ of size $N_c \times N_c$, where $N_c$ is the number of distinct concepts present in the annotations of $d$. This matrix is initialized as follows:

\begin{itemize}
\item $M_{ij}^d=M{ji}^d=1$, if $c_i$ and $c_j$ co-occur in a same sentence of the document $d$,
\item $M_{ij}^d=0.5$ and $M_{ji}^d=1$, if $c_i \preceq c_j$ in the reference ontology $O$, 
\item $M_{ii}^d=1$,
\item otherwise $M_{ij}^d=0$ with $i \neq j$ .
\end{itemize}

The affinity matrix can be used in several existing graph-based algorithms that aim to rank  the nodes according to the neighbors contributions. We have chosen the
regularization framework proposed in \cite{Zhou2004}, which can be summarized with the following formula:

\begin{eqnarray} 
   R^d = ((1 - \alpha) \cdot (I - \alpha S^d)^{-1} \cdot Y^T)^T 
\end{eqnarray}

Here, $R$ is the vector representing the rank of concepts. This is obtained by finding an optimal smoothed function that best fits a given vector $Y$,
which is achieved by applying the laplacian operator over the affinity matrix $M^d$ as follows:

\[ S^d = D^{-1/2} \cdot M^d \cdot D^{-1/2} \]

In our case, the vector $Y$ consists of the frequencies of each concept in the document $d$. The parameter $\alpha$ is directly related
to the smoothness of the approximation function (we set it to $\alpha=0.9$). 

An alternative to this method is to use a centrality-based algorithm over $M^d$. Our preliminary experiments over the HeC collections showed that this method
obtains very similar ranks to the previous one. 

Once the rank $R^d$ is obtained, the normalization process consists in selecting the top-scored concepts of each dimension to represent the $d$'s fact.

As an example, the multidimensional fact resulted from the document presented in Figure \ref{taggedtext} is as follows: \\

{\small 
\begin{tabular}{l}
{\tt ( ResearchActivity:C1709323, PopulationGroup:C0007457, AgeGroup:C0008059, }\\
{\tt   Disease:C1384600, ImmunologyFactor:C0063717, ...) }\\
\end{tabular}
}

\subsection{Building 3D conceptual maps}

As mentioned in the introduction, our main aim is to build a browseable representation of the semantic spaces defined in the previous section.
For this purpose, we define the 3D conceptual map, which is a sequence of different layers that correspond to different dimensions 
expressed at some detail level (category). In this map, concepts are visualized as balls, which are placed within  their corresponding layer 
with a size proportional to their relevance w.r.t. the target collection. Concept bridges (or conceptual associations) are visualized as 
links between concepts of adjacent layers. 3D maps are built from the normalized conceptual representation described in the previous section,
by using a series of basic operations, which are described in turn.

\subsubsection{Basic operations}
The basic operations that can be defined over a dimension $D_i$ are the following ones:

\begin{description}
\item[Layer definition,] which establishes the concepts that will be placed at the layer. 
This operation can be done either by specifying one dimension category or through a 
keyword-based query. In the first case, all the concepts of the dimension
category are visualized, whereas in the second case only the most specific concepts in $D_i$ 
whose lexical forms match the query are visualized.

\item[Concept containment,] returns all the sub-concepts of a selected concept $q$ of a dimension $D_i$. Formally, 

\[ descendants(D_i, q) = \{c \; | \; c \in D_i \;\wedge\; c \preceq q \} \]

\item[Text containment,] which returns true if there exists some concept $c \preceq q $ whose lexicon, $lex(c)$, 
matches the specified keywords:

\[ contains(D_i, q, kywds) )= \{c | c \in D_i \;\wedge\; c \preceq q \;\wedge\; matches(lex(c),kywds) \} \] 

\item[Direct subconcepts,] denoted $children(D_i,c)$, which returns the set of direct sub-concepts of a concept $C$.
This operation is used to browse the taxonomy downwards (drill-down operation).

\end{description}

All these operations are efficiently performed by using the interval algebra over the $\mathcal{L}^-$ scheme associated to the
ontology concepts.

\subsubsection{Aggregations}
Summarization is one of the main purposes of the proposed analytical tool to facilitate the exploration of the collection contents.
Similarly to OLAP-like systems, summarization is performed through well-defined aggregations over the semantic annotations of the
objects collections. More specifically, the following aggregations are performed to visualize summarized information:

\begin{description}
\item[Concept Relevance.] The relevance of a concept $c$ is calculated by aggregating the relevance of its sub-concepts w.r.t each specific collection. 
Formally, 

\[ Rel_{{\mathcal Col}}(c,D_i) = \Gamma_{\forall c' \in descendants(D_i, c)} \; score_{\mathcal Col}(c') \] 

where $\Gamma$ is an aggregation function (e.g., sum, avg, and so on) and $score$ is the function that is evaluated against the collection. The simplest
scoring function is the number of hits, namely: 

\[ score_{\mathcal Col}(c) = hits_{\mathcal Col}(c')=count(\{ d | d \in {\mathcal Col}, fact(d)[D_i]=c'\})\]

Alternatively, the scoring function can take into account the relevance of each concept in the documents it appears. Thus, we can aggregate the relevance
scores estimated to select concept facts (see Formula 1) as follows:

\[ score_{\mathcal Col}(c) = \sum_{d \in {\mathcal Col}, \exists i, fact(d)[D_i]=c} { R^d[c] } \]

\item[Concept Associations.] Given two dimension levels $L^i_n$ and $L^j_m$, belonging to dimensions $D_i$ and $D_j$ ($i \neq j$) respectively, the 
following 2D cube stores the aggregated contingency tables necessary for correlation analysis:

\[ CUBE_{{\mathcal Col}}(L^i_n,L^j_m)=\{ (c_i,c_j, n_{i,j}, n_i, n_j ) | c_i \in  L^i_n \wedge c_j \in L^j_m \} \]

Here $n_{i,j}$ measures the number of objects in the collection where $c_i$ and $c_j$ co-occur, $n_i$ is the number of objects where $c_i$ occurs, and 
$n_j$ is the number of objects where $c_j$ occurs. Notice that $n_i$ and $n_j$ are calculated in a similar way  as concept relevance. The contingency table for
 each pair $(c_i, c_j)$ is calculated as shown in Table \ref{contingency}.

\begin{table}
\begin{center}
\begin{tabular}{|l|l|l|}
\hline
           & {\bf $c_i$ }           & {\bf $\bar{c_i}$ }\\ \hline
{\bf $c_j$ }        & $n_{i,j}$       & $n_j - n_{i,j}$ \\ \hline
{\bf $\bar{c_j}$ }  & $n_i - n_{i,j}$ & $N_{Col} - n_i - n_j$\\ \hline
\end{tabular}
\end{center}
\caption{Contingency table for scoring bridges}
\label{contingency}
\end{table}

The measures $n_{i,j}$, $n_i$ and $n_j$ are calculated as follows:

\begin{eqnarray*}
n_{i,j} & =  |\{d| d \in {\mathcal Col} \wedge fact(d)[D_i] \preceq c_i \wedge fact(d)[D_j] \preceq c_j \}| \\
n_{i}   & =  |\{d| d \in {\mathcal Col} \wedge fact(d)[D_i] \preceq c_i \}| \\
n_{j}   & =  |\{d| d \in {\mathcal Col} \wedge fact(d)[D_j] \preceq c_j \}|
\end{eqnarray*}

\item[Semantic Bridges.] A semantic bridge is a strong association between concepts which has  good evidence in the target collection. Bridges are calculated
from contingency tables by defining a scoring function $\phi(c_i,c_j)$. In this way, bridges will be those concept associations whose score is greater than a specified 
threshold $\delta$:

\[ Bridges_{{\mathcal Col}}^\phi(L_i, L_j) =  \{(c_i,c_j,\phi(c_i,c_j)) | \phi(c_i,c_j) > \delta \} \]

As an example, we can use the interest factor as score, that is:

\[\phi(c_i,c_j)= \frac{n_{i,j} \cdot N}{n_i \cdot n_j}\]

In our current setting, we use a series of well-known interestingness measures such as log likelihood ratio, mutual information, interest factor and F1-measure.
%Explain $\delta$-maximum interestingness pairs. (?)

\end{description}

\subsubsection{Browsing conceptual maps}
Two main browsing operations can be performed in a conceptual map: (1) expand a concept into its sub-concepts, and (2) go to a ranked list of objects 
associated to the clicked map elements (concepts and bridges). The semantics of these operations corresponds to the well-known drill-down and 
drill-through OLAP operations.

\begin{description}
\item[Drill-down:] If we expand a concept $c$ in the 3D map, it must be updated accordingly. Thus, the concept $c$ is substituted by its children in the $O$'s taxonomy,
bridges involved by $c$ are removed from the map, and new bridges are calculated for the sub-concepts of $c$ and drawn in the map. 

\item[Drill-through:] If a concept (bridge) is selected for drill-through, the system must retrieve the objects of the target collection relevant to it.
The ranked list of objects is shown in a separate list (e.g., tab) ordered by relevance. Notice that we can simply use the score calculated to
construct facts (i.e., $R^d$) for ranking documents w.r.t. concepts, formally:

\[ relevance(d, c) = R^d[c] \]

For ranking documents w.r.t. bridges, we just combine the scores of the involved concepts in the selected bridge: 

\[ relevance(d,(c_i,c_j,\phi))=relevance(d, c_i)\cdot relevance(d, c_j) \]

\end{description}

\section{Conclusions}
In this paper we have presented a novel semantics-aware integration and visualization paradigm
that allows users to easily explore and navigate discovered relations between data and
web resources. The contribution is two-fold. On one hand, we provide the infrastructure for
integrating different information resources through semantic annotation with domain ontologies.
On the other hand, users can interactively build conceptual maps according to their
requirements and explore them with classical OLAP-style operations such as roll-up and drill-down.
Some future work includes the refinement of the created dimension hierarchies in order
to account for more meaningful aggregations and also to devise more efficient calculation of new bridges.
Finally, we plan to develop an on-line service to provide conceptual maps on-demand.

%\nocite{*}
%GATHER{references.bib}
\bibliographystyle{splncs}
\bibliography{references}

\end{document}